\documentclass[%
  a4paper, twocolumn, pra, superscriptaddress, amsmath, amssymb, floatfix
]{revtex4-1}
\usepackage{graphicx}
\usepackage{epstopdf}
\usepackage[table]{xcolor}
\usepackage{braket}
\usepackage{bbm}
\usepackage{siunitx}
\usepackage[colorlinks,bookmarksopen=false]{hyperref}
\usepackage{natbib}
\hypersetup{%
  pdfstartview={FitH},
  linkcolor=blue,
  citecolor=blue,
  filecolor=black,
  menucolor=black,
  urlcolor =blue
}

\newcommand{\im}{\ensuremath{\textup{i}}}

\newcommand{\rp}{\ensuremath{\mathfrak{Re}}}
\newcommand{\ip}{\ensuremath{\mathfrak{Im}}}

\newcommand{\op}[1]{\ensuremath{\mathsf{\hat{#1}}}}
\newcommand{\lop}[1]{\ensuremath{\mathcal{#1}}}
\newcommand{\vecop}[1]{\ensuremath{\boldsymbol{\mathsf{\hat{#1}}}}}
\newcommand{\uop}{\op{\mathbbm{1}}}

\newcommand{\opnorm}[1]{\left\lVert#1\right\rVert}

\newcommand{\pulse}{\mathcal{E}}

\newcommand{\myvec}[1]{\boldsymbol{#1}}

\newcommand{\dd}{\ensuremath{\mathrm{d}}}

\begin{document}

\title{%
  Quantum Optimal Control for Mixed State Squeezing in Cavity Optomechanics
}

\author{Daniel Basilewitsch}
\author{Christiane P. Koch}
\email{christiane.koch@uni-kassel.de}
\author{Daniel M. Reich}
\affiliation{Theoretische Physik, Universit\"{a}t Kassel, D-34132 Kassel,
Germany}

\date{\today}

\begin{abstract}
  The performance of key tasks in quantum technology, such as accurate state
  preparation, can be maximized by utilizing external controls and deriving
  their shape with optimal control theory. For non-pure target states, the
  performance measure needs to match both angle and length of the generalized
  Bloch vector. We introduce a measure based on this simple geometric picture
  that separates angle and length mismatch into individual terms and apply the
  ensuing optimization framework to maximize squeezing of an optomechanical
  oscillator at finite temperature. Our results show that shaping the cavity
  drives can speed up squeezed state preparation by more than two orders of
  magnitude. Cooperativities and pulse shapes required to this end are fully
  compatible with current experimental technology.
\end{abstract}

\maketitle

\section{Introduction} \label{sec:intro}
The ability to precisely control quantum systems is a prerequisite for
harnessing quantum effects for quantum technology. Quantum optimal control
provides a set of tools for deriving protocols to implement key tasks such as
the preparation of non-classical states or the generation of
entanglement~\cite{Glaser.EPJD.69.279}. This approach can be used to determine,
for example, the minimum time to carry out a given task with desired accuracy,
even if the dynamics is not amenable to an analytical
solution~\cite{CanevaPRL09, GoerzJPB11, sorensen2016exploring}. Recent examples
include the fast and accurate preparation of a circular state, i.e., a Rydberg
state with maximum projection angular momentum quantum number, for quantum
sensing~\cite{PatschPRA18}, and the determination of the fastest universal set
of gates for quantum computing with superconducting transmon
qubits~\cite{GoerzNPJQI17}.

Fast control protocols are particularly important for open quantum systems for
which the interaction with the environment cannot be
neglected~\cite{Breuer.book}. An obvious control strategy is to `beat' the
decoherence resulting from the interaction with the environment. For quantum
systems with Markovian, i.e., memoryless dynamics, this is often the only
option~\cite{Koch.JPhysCondensMatter.28.213001}. In contrast, strongly coupled
environmental modes giving rise to significant system-environment correlations
and non-Markovian dynamics are not necessarily detrimental but can also be
exploited for control~\cite{BasilewitschNJP17, ReichSciRep15}.

An alternative approach to controlling open quantum systems consists in
engineering driven dissipative dynamics in such a way that the desired target
state becomes the steady state of the ensuing open system
evolution~\cite{PoyatosPRL96}. This approach is particularly promising when the
timescale of decoherence is comparable to or faster than that of the coherent
evolution. Driven dissipative evolution is inherently robust against noise---a
rather favorable feature in view of experimental implementation. In this
context, preparation of non-classical states~\cite{krauter2011entanglement,
lin2013dissipative, shankar2013autonomously, kienzler2014quantum,
kimchi2016stabilizing} and generation of non-equilibrium quantum
phases~\cite{syassen2008strong, schindler2013quantum} have successfully been
demonstrated with trapped atoms and ions as well as superconducting qubits. For
the example of trapped ions, determining the key parameters of the driven
dissipative dynamics by quantum optimal control is predicted to allow reaching
the fundamental performance limits~\cite{HornNJP18}.

Another platform ideally suited for implementing driven dissipative dynamics is
cavity optomechanics~\cite{Aspelmeyer.RevModPhys.86.1391,
CavityOptomechanics.book}, where a mechanical resonator is coupled to an optical
or microwave cavity. Optomechanical systems are promising candidates for
quantum-enhanced sensing, coherent light-matter interfaces, and fundamental
tests of quantum mechanics. In particular, the cavity drive can be employed to
generate arbitrary quantum states of the mechanical
oscillator~\cite{Aspelmeyer.RevModPhys.86.1391}, including strongly squeezed
states. These states are useful in applications such as quantum information
processing~\cite{Yonezawa2010}, atomic clocks~\cite{Leroux.PRL.104.250801} or,
most prominently, quantum-enhanced sensing~\cite{Schnabel20171, Hosten2016}
where they allow to increase sensitivity of e.g.\ gravitational wave
detectors~\cite{Aasi2013, Grote.PRL.110.181101}. Squeezed states can be
generated in various physical platforms~\cite{Andersen.PhysScr.91.053001}. In
cavity optomechanics, driven-dissipative evolution can be used to produce
substantially squeezed states~\cite{Kronwald.PRA.88.063833,
Wollman.Science.349.952}. While preparation of the mechanical resonator in
a pure quantum state remains a major goal of cavity optomechanics, interesting
non-classical states can be realized also at finite
temperature~\cite{Palomaki710, Wollman.Science.349.952,
Pirkkalainen.PhysRevLett.115.243601}. This is true in particular for squeezed
states since there is a trade off between squeezing strength and purity.

Cavity drives for state preparation in cavity optomechanics are typically taken
to have constant amplitude~\cite{Aspelmeyer.RevModPhys.86.1391}. On the other
hand, pulsed excitation has recently been shown to allow for probing the
resonator state with minimal heating~\cite{Meenehan.PhysRevX.5.041002}. This
raises the question whether explicitly time-dependent amplitudes of the cavity
drives can also be used to improve state preparation protocols. Here, we
specifically ask by how much, at a given non-zero temperature, suitable pulse
shapes can speed up the preparation of the mechanical oscillator in a squeezed
state.

To derive the pulse shape of the cavity drives, we employ optimal control theory
which needs to target a mixed steady state, due to finite temperature. Standard
optimization functionals, defined originally for pure target states, cannot be
used in this case; and alternative formulations using, e.g., the Hilbert-Schmidt
distance need to be employed~\cite{Xu.JChemPhys.120.6600}. We give an intuitive,
geometrical explanation for the failure of the standard functionals by
visualizing the dynamics on the generalized Bloch sphere. This picture is also
useful to elucidate the requirements an optimization functional targeting mixed
states need to fulfill. Using this geometric interpretation, we furthermore
refine the functional based on the Hilbert-Schmidt
distance~\cite{Xu.JChemPhys.120.6600} to one that seeks to match the target
state's Bloch vector angle and length separately. We employ both functionals to
optimize the preparation of a mechanical resonator in a squeezed state and
compare their performance.

The paper is organized as follows. The framework of quantum optimal control
theory is presented in Sec.~\ref{sec:oct}. In Sec.~\ref{subsec:Krotov} we
briefly review Krotov's method~\cite{Krotov.book}, our optimization algorithm of
choice, in Sec.~\ref{subsec:funcs_old} we illustrate the failure of standard
functionals, and in Sec.~\ref{subsec:funcs_new} we construct target functionals
based on Bloch vector angle and length. Section~\ref{sec:app} is dedicated to
the application of this methodology to preparing a squeezed state at
finite temperature in cavity optomechanics. We introduce the model and control
problem in Sec.~\ref{subsec:model}, present our results in
Sec.~\ref{subsec:results} and discuss the performance of various target
functionals in Sec.~\ref{subsec:results:perf}. Finally, Sec.~\ref{sec:con}
concludes.

\section{Quantum Optimal Control Theory} \label{sec:oct}
Quantum control assumes that the dynamics of a quantum system can be steered,
typically by a set of external driving fields $\left\{\pulse_{k}(t)\right\}$.
How to choose the external drives in the best possible way is the subject of
quantum optimal control theory (OCT)~\cite{Glaser.EPJD.69.279}. Optimality is
sought using an optimization functional,
\begin{equation} \label{eq:oct:J_tot}
  \begin{aligned}
    J\left[\left\{\pulse_{k}\right\}, \left\{\op{\rho}_{l}\right\}\right]
    &=
    D\left[\left\{\op{\rho}_{l}(T)\right\}\right]
    \\
    &\quad+
    \int_{0}^{T} \dd t\, g\left[\left\{\pulse_{k}(t)\right\},
    \left\{\op{\rho}_{l}(t)\right\}, t\right],
  \end{aligned}
\end{equation}
where $\left\{\op{\rho}_{l}(t)\right\}$ is a set of forward propagated states,
$D\left[\left\{\op{\rho}_{l}(T)\right\}\right]$ is the figure or merit at final
time $T$ and $g$ captures additional costs or constraints at intermediate times,
for instance by restricting the field spectra or by penalizing population in
certain subspaces.

A control problem is tackled by choosing (i) an appropriate optimization
functional $J$ and (ii) an appropriate method to minimize $J$. While $J$
captures the physics of the problem, the choice of the method is relevant as
well, since it often determines whether a solution can be found in practice.
Gradient-based techniques hold the promise of faster convergence than
gradient-free methods~\cite{Glaser.EPJD.69.279}. However, they require the
ability to determine the derivatives of the functional with respect to the
states. Whether this is feasible or not depends on the choice of the
functional. In the following, we focus on Krotov's method~\cite{Krotov.book,
Konnov.AutomRemContr.60.1427}, a gradient-based algorithm, but our
considerations are valid for any gradient-based method requiring functional
derivatives.

The general idea of any gradient-based method is to find an extremum of the
total functional~\eqref{eq:oct:J_tot} using gradient information with respect to
changes in the control fields $\{\pulse_{k}(t)\}$. The extremum condition
together with the requirement to fulfill the dynamical equations leads to two
equations of motion - one for the states $\left\{\op{\rho}_{l}(t)\right\}$ and
one for the so-called co-states $\{\op{\chi}_{l}(t)\}$. While the former
corresponds to the usual forward propagation in time starting at the initial
condition $\left\{\op{\rho}_{l}(0)\right\}$, the latter can be interpreted as
a backward propagation in time, starting at $\{\op{\chi}_{l}(T)\}$. This
``initial'' condition for the backward propagation contains information about
the physical final time target encoded by the functional $D$. The optimization
algorithms then tries to match both forward and backward propagated states, thus
ensuring approach towards the target $\left\{\op{\rho}_{l}(T)\right\}$ at final
time, which in turn minimizes functional~\eqref{eq:oct:J_tot}.

\subsection{Gradient-based optimization with Krotov's method}
\label{subsec:Krotov}

Krotov's method~\cite{Krotov.book, Konnov.AutomRemContr.60.1427} is
a sequential optimization technique with built-in monotonic convergence.
A possible choice of the functional $g$ is~\cite{Palao.PRA.68.062308}
\begin{align} \label{eq:krotov:g}
  g\left[\left\{\pulse_{k}(t)\right\}\right]
  =
  \sum_{k} \frac{\lambda_{k}}{S_{k}(t)} \left(\pulse_{k}(t)
  - \pulse_{k}^{\mathrm{ref}}(t)\right)^{2},
\end{align}
where $\pulse_{k}^{\mathrm{ref}}(t)$ is a reference field (taken to be the field
from the last iteration), $S_{k}(t) \in (0,1]$ a shape function to smoothly
switch $\pulse_{k}(t)$ on and off, and $\lambda_{k}$ a parameter that controls
the step size. Using Eq.~\eqref{eq:krotov:g}, the update equation for the field
$\pulse_{k}(t)$ becomes~\cite{Reich.JChemPhys.136.104103, GoerzNJP14}
\begin{widetext}
  \begin{align} \label{eq:krotov:update}
    \pulse_{k}^{(i+1)}(t)
    =
    \pulse_{k}^{\mathrm{ref}}(t)
    +
    \frac{S_{k}(t)}{\lambda_{k}} \ip\left\{%
      \sum_{l} \Braket{%
        \op{\chi}^{(i)}_{l}(t)\;,\;
        \frac{\partial \lop{L}\left[\left\{\pulse_{k'}\right\}\right]}{\partial
        \pulse_{k}} \Big|_{\{\pulse^{(i+1)}_{k'}(t)\}}
        \op{\rho}^{(i+1)}_{l}(t)
      }
    \right\},
  \end{align}
\end{widetext}
where $\braket{\op{A}, \op{B}} \equiv \mathrm{Tr}\{\op{A}^{\dagger} \op{B}\}$ is
the Hilbert-Schmidt overlap and $\lop{L}\left[\left\{\pulse_{k}\right\}\right]$
the field-dependent generator of the dynamics, e.g.\ the Liouvillian of
a Lindblad master equation~\cite{Breuer.book}. $\{\op{\rho}^{(i+1)}_{l}(t)\}$
are forward propagated states,
\begin{subequations}
  \begin{equation}
\label{eq:krotov:fw_states}
    \frac{\dd}{\dd t} \op{\rho}^{(i+1)}_{l}(t)
    =
    - \frac{\im}{\hbar} \lop{L}\left[\left\{\pulse^{(i+1)}_{k}\right\}\right]
    \op{\rho}^{(i+1)}_{l}(t)\,,
  \end{equation}
with initial conditions
\begin{equation}
   \label{eq:krotov:fw_states_bound}
    \op{\rho}^{(i+1)}_{l}(0)
    =
    \op{\rho}_{l}(0)\,,
\end{equation}
\end{subequations}
whereas the co-states $\{\op{\chi}^{(i)}_{l}(t)\}$ are solutions of the adjoint
equation of motion,
\begin{subequations} \label{eq:krotov:co}
  \begin{equation}
    \label{eq:krotov:co_states}
    \frac{\dd}{\dd t} \op{\chi}^{(i)}_{l}(t)
    =
    \frac{\im}{\hbar}
    \lop{L}^{\dagger}\left[\left\{\pulse^{(i)}_{k}\right\}\right]
    \op{\chi}^{(i)}_{l}(t)\,,
  \end{equation}
and their `initial' condition (at final time $T$) is determined by the target
functional,
\begin{equation}
     \label{eq:krotov:co_states_bound}
    \op{\chi}^{(i)}_{l}(T)
    =
    - \nabla_{\op{\rho}_{l}(T)} D \big|_{\{\op{\rho}^{(i)}_{l'}(T)\}}\,.
\end{equation}
\end{subequations}
The superscripts $(i+1)$ and $(i)$ in
Eqs.~\eqref{eq:krotov:update}-\eqref{eq:krotov:co} indicate the current and
previous iteration in the optimization procedure, respectively. The choice of
$D\left[\left\{\op{\rho}_{l}(T)\right\}\right]$ enters via
Eq.~\eqref{eq:krotov:co_states_bound}, where the derivative of functional $D$
with respect to $\op{\rho}_{l}(T)$ needs to be evaluated. For a detailed
derivation of Krotov's method in the context of quantum control see
Ref.~\cite{Reich.JChemPhys.136.104103}.

\subsection{Failure of overlap-based functionals for mixed state targets}
\label{subsec:funcs_old}
State transfers, where a set of initial states $\left\{\op{\rho}_{l}(0)\right\}$
must simultaneously be transferred into a set of target states
$\left\{\op{\rho}_{l}^{\mathrm{trg}}\right\}$, represent a standard control
problem. Solving this problem requires a reliable measure $D(\op{\rho}_{l}(T),
\op{\rho}_{l}^{\mathrm{trg}})$ for the state distance between forward propagated
state $\op{\rho}_{l}(T)$ and target $\op{\rho}_{l}^{\mathrm{trg}}$. For two pure
states $\Psi_{1}, \Psi_{2}$, a standard choice is~\cite{Palao.PRA.68.062308}
\begin{subequations} \label{eq:oct:F_re_sm}
  \begin{align}
    D_{\mathrm{re}}\left(\Psi_{1}, \Psi_{2}\right)
    &=
    1 - \rp\left\{\tau\right\},
    \quad \text{or}
    \\
    D_{\mathrm{sm}}\left(\Psi_{1}, \Psi_{2}\right)
    &=
    1 - \left|\tau\right|^{2},
  \end{align}
\end{subequations}
where $\tau = \Braket{\Psi_{1} | \Psi_{2}} \in \mathbb{C}$, $|\tau| \leq 1$, is
the complex overlap in Hilbert space. Both functionals rely on $\tau$ to serve
as a distance measure in state space, and any OCT algorithm that minimizes
$D_{\mathrm{re}}$ or $D_{\mathrm{sm}}$ necessarily maximizes
$\rp\left\{\tau\right\}$ or $|\tau|$, respectively.
Equations~\eqref{eq:oct:F_re_sm} cannot, however, simply be generalized to
non-pure states. For density matrices $\op{\rho}_{1}$, $\op{\rho}_{2}$, the
overlap in Liouville space, $\tau = \braket{\op{\rho}_{1},\op{\rho}_{2}} \in
\mathbb{R}$, defined in terms of the Hilbert-Schmidt overlap, becomes real, and
minimizing $D_{\mathrm{re}}$, $D_{\mathrm{sm}}$ implies maximizing $\tau$.
Unfortunately, $\tau$ is no longer a reliable measure for closeness once
$\op{\rho}_{1}, \op{\rho}_{2}$ are \emph{both} mixed.

We illustrate the problem with the simplest example of a quantum system,
a qubit. Representing the qubit state in the canonical basis,
$\left\{\Ket{0},\Ket{1}\right\}$, consider
\begin{align} \label{eq:oct:qubit_states}
  \op{\rho}(\alpha)
  =
  \begin{pmatrix}
    \alpha & 0 \\ 0 & 1-\alpha
  \end{pmatrix}\,,
  \qquad
  \op{\rho}^{\mathrm{trg}}
  =
  \begin{pmatrix}
    \beta & 0 \\ 0 & 1-\beta
  \end{pmatrix}\,,
\end{align}
where $0 \leq \alpha$, $\beta \leq 1$. Both states are equivalent iff $\alpha
= \beta$ and $\tau_{\mathrm{trg}} \equiv \Braket{\op{\rho}(\beta),
\op{\rho}^{\mathrm{trg}}} = \beta^{2} + (1-\beta)^{2}$. However, for pure states
such as $\op{\rho}(1)$ or $\op{\rho}(0)$, $\tau=\beta$ or $\tau=1-\beta$. This
results in $\tau > \tau_{\mathrm{trg}}$ for $\beta \in (\frac{1}{2}, 1)$ or
$\beta \in (0, \frac{1}{2})$, respectively. Thus, pure states maximize $\tau$
and thus the figure of merit~\eqref{eq:oct:F_re_sm} despite being obviously
different from the target $\op{\rho}^{\mathrm{trg}}$. Moreover, for the
completely mixed state $\op{\rho}^{\mathrm{trg}} = \mathrm{diag}\{0.5, 0.5\}$,
we find $\tau = \braket{\op{\rho}(\alpha), \op{\rho}^{\mathrm{trg}}}
= \frac{1}{2}$ for all $\alpha$. In this case, $\tau$ is not even able to
indicate differences at all.

\begin{figure}[tb]
  \centering
  \includegraphics{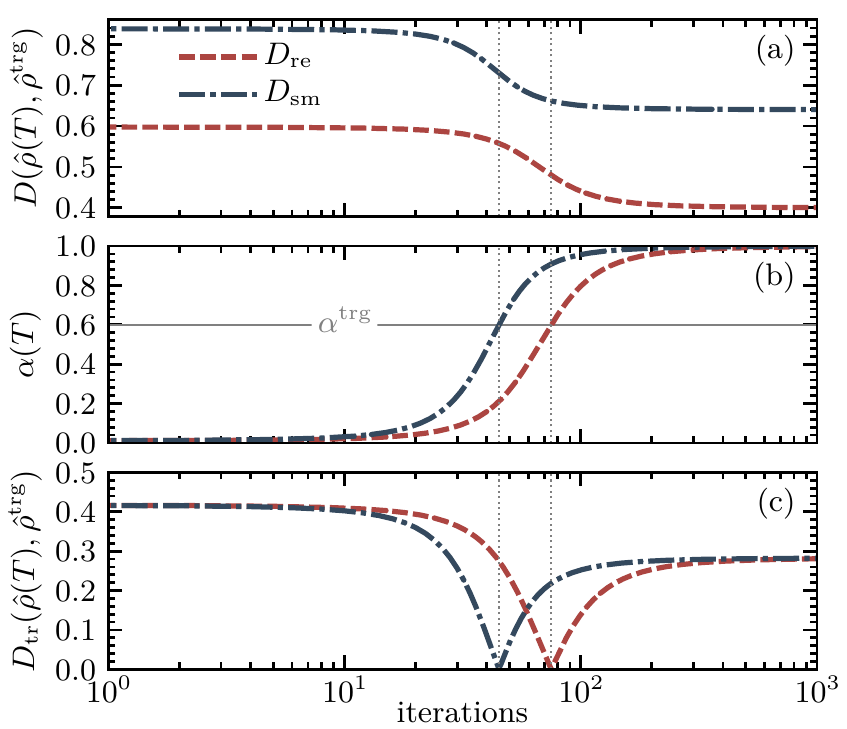}
  \caption{%
    Optimization results for a qubit, where the dynamics is governed by
    Eq.~\eqref{eq:oct:LvN_toy}. The initial state is $\op{\rho}(0)
    = \Ket{1}\Bra{1}$, the target state $\op{\rho}^{\mathrm{trg}}
    = \mathrm{diag}\{0.6, 0.4\}$ and the initial
    field $u(t) = 0.01$ with total propagation time $T=1$.
    (a) Final-time functionals $D_{\mathrm{re}}$ and $D_{\mathrm{sm}}$, cf.
    Eq.~\eqref{eq:oct:F_re_sm}, as a function of the number of iterations.
    (b) Population $\alpha(T)$ in $\Ket{0}$ at final time. The horizontal line
    indicates the respective population of the target state
    $\op{\rho}^{\mathrm{trg}}$.
    (c) Trace distance $D_{\mathrm{tr}}$, cf. Eq.~\eqref{eq:tr_dist}, between
    propagated state $\op{\rho}(T)$ and target state $\op{\rho}^{\mathrm{trg}}$.
  }
  \label{fig:toy}
\end{figure}

The ill-definedness of the overlap-based functionals~\eqref{eq:oct:F_re_sm} in
case of mixed target states is easily demonstrated by a toy control problem.
Consider a qubit whose dynamics is determined by a purely dissipative master
equation~\cite{Breuer.book},
\begin{equation} \label{eq:oct:LvN_toy}
  \begin{aligned}
    \im \hbar \frac{\dd}{\dd t} \op{\rho}(t)
    &=
    \lop{L}\left[u(t)\right] \op{\rho}(t)
    \\
    &=
    \im \hbar u(t) \left[
      \op{\sigma}_{-} \op{\rho}(t) \op{\sigma}_{+}
      - \frac{1}{2} \left\{%
        \op{\sigma}_{+} \op{\sigma}_{-}, \op{\rho}(t)
      \right\}
    \right],
  \end{aligned}
\end{equation}
where $\op{\sigma}_{-} (\op{\sigma}_{+})$ are the standard lowering (raising)
operators and $u(t) \geq 0$ is a time-dependent, controllable decay rate. We
choose the initial state of the qubit to be $\op{\rho}(0) = \Ket{1}\Bra{1}$.
Thus, we can reach any diagonal state $\op{\rho}(T) = \alpha(T)\Ket{0}\Bra{0}
+ (1-\alpha(T))\Ket{1}\Bra{1}$ with $\alpha(T)>0$, since $\alpha(T)$ can be
controlled by appropriately choosing $u(t)$ for $t \in [0, T]$.
Figure~\ref{fig:toy} presents optimization results for a mixed state target,
$\op{\rho}^{\mathrm{trg}} = 0.6 \Ket{0}\Bra{0} + 0.4 \Ket{1}\Bra{1}$, employing
the functionals~\eqref{eq:oct:F_re_sm}. Figure~\ref{fig:toy}(a) shows the
monotonic decrease of both functionals over the number of iterations, as
expected for Krotov's method, while Fig.~\ref{fig:toy}(b) plots the
corresponding final time ground state population $\alpha(T)$. The optimization
starts with a fairly low ground state population, $\alpha(T) \sim 0$, due to the
non-optimal, i.e., too small, guess field $u(t)$. The decay rate is increased
during optimization such that $\alpha(T) \sim 1$ after convergence is reached.
This result maximizes the overlap since $\tau_{\mathrm{opt}} \equiv
\Braket{\op{\rho}_{\mathrm{opt}}(T), \op{\rho}^{\mathrm{trg}}}
> \Braket{\op{\rho}^{\mathrm{trg}}, \op{\rho}^{\mathrm{trg}}} \equiv
\tau_{\mathrm{eq}}$ with $\op{\rho}_{\mathrm{opt}}(T) = \Ket{0}\Bra{0}$, and
thus realizes smaller values of the functionals $D_{\mathrm{re}}$ and
$D_{\mathrm{sm}}$. However, this is not what the optimization is supposed to
achieve. Figure~\ref{fig:toy}(c) shows the trace distance $D_{\mathrm{tr}}$ (a
reliable measure for the closeness of states, as we will discuss in
Sec.~\ref{subsec:funcs_new}) between $\op{\rho}(T)$ and
$\op{\rho}^{\mathrm{trg}}$ as a function of the number of iterations. A minimum
is observed at the correct value $\alpha(T) = \alpha^{\mathrm{trg}} = 0.6$. The
increase of $D_{\mathrm{tr}}$ as the iterative algorithm proceeds, which is due
to further minimization of $D_{\mathrm{re}}$ and $D_{\mathrm{sm}}$, illustrates
that the optimization misses the desired target.

This observation can be fully generalized to $N$-level systems with a simple
geometric picture~\cite{Bertlmann.JPhysAMathTheor.41.235303}. By choosing
a basis of traceless, Hermitian $N \times N$ matrices, $\{\op{A}_{i}\}$, with
$\braket{\op{A}_{i}, \op{A}_{j}} = \delta_{i,j}$, we can write $\op{\rho}$ as
\begin{align}
  \op{\rho}
  =
  \frac{1}{N} \uop_{N} + \myvec{r} \cdot \vecop{A}\,,
\end{align}
where $\myvec{r} = (a_{1}, a_{2}, \dots)^{\top}$ is the generalized Bloch
vector, containing the expansion coefficients for matrices $\vecop{A}
= (\op{A}_{1}, \op{A}_{2}, \dots)^{\top}$. Then, the Hilbert-Schmidt overlap of
two states $\op{\rho}_{1}, \op{\rho}_{2}$ becomes
\begin{align} \label{eq:tau_bloch}
  \tau
  =
  \Braket{\op{\rho}_{1}, \op{\rho}_{2}}
  =
  \frac{1}{N} + \myvec{r}_{1} \cdot \myvec{r}_{2}
  =
  \frac{1}{N} + \left|\myvec{r}_{1}\right| \left|\myvec{r}_{2}\right|
  \cos(\theta)\,,
\end{align}
where $|\myvec{r}_{1}|, |\myvec{r}_{2}|$ are the lengths of the respective Bloch
vectors and $\theta$ is the angle between them. Hence, maximization of $\tau$
means maximizing $|\myvec{r}_{1}|, |\myvec{r}_{2}|$ and minimizing $\theta$.

\begin{figure}[tb]
  \centering
  \includegraphics{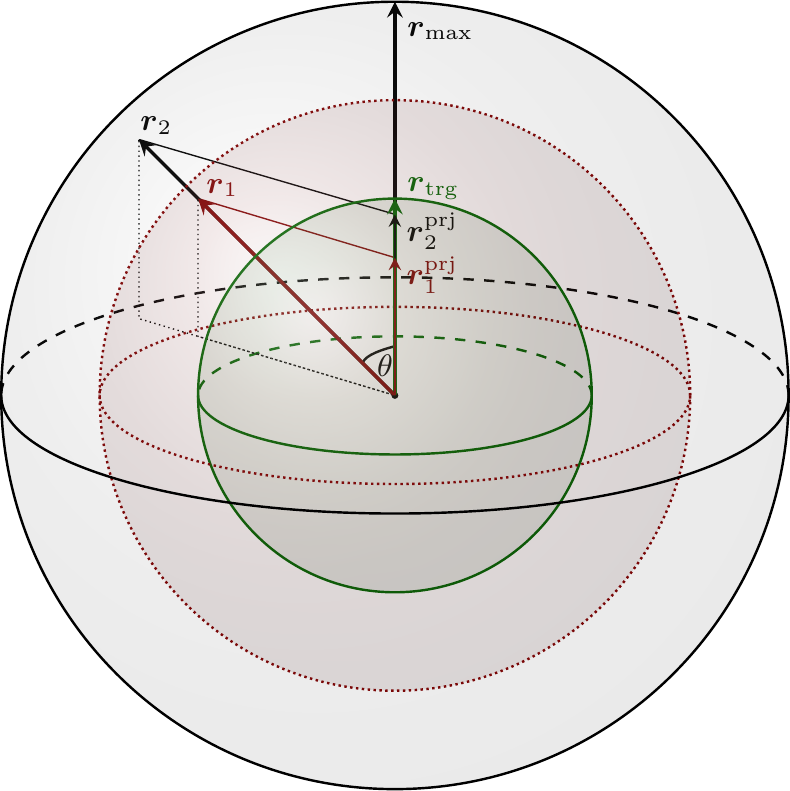}
  \caption{%
    Bloch vectors $\myvec{r}_{1}, \myvec{r}_{2}, \myvec{r}_{\mathrm{trg}},
    \myvec{r}_{\mathrm{max}}$ in a generalized Bloch sphere. The outer sphere
    indicates pure states while the inner spheres correspond to mixed states.
    $\myvec{r}_{1}^{\mathrm{prj}}, \myvec{r}_{2}^{\mathrm{prj}}$ are the
    projections of $\myvec{r}_{1}, \myvec{r}_{2}$ onto
    $\myvec{r}_{\mathrm{trg}}$.
  }
  \label{fig:sphere}
\end{figure}

Figure~\ref{fig:sphere} illustrates the behavior of the overlap geometrically,
comparing two states $\op{\rho}_{1}, \op{\rho}_{2}$ with Bloch vectors
$\myvec{r}_{1}, \myvec{r}_{2}$ to the target state $\op{\rho}_{\mathrm{trg}}$
with $\myvec{r}_{\mathrm{trg}}$. Here, we assume $\myvec{r}_{1} \parallel
\myvec{r}_{2}$ and $|\myvec{r}_{2}| > |\myvec{r}_{1}|
> |\myvec{r}_{\mathrm{trg}}|$. In this case, the angles $\theta_1$, $\theta_2$
of $\myvec{r}_{1}$ and $\myvec{r}_{2}$ with $\myvec{r}_{\mathrm{trg}}$ are
identical but the purer state $\myvec{r}_{2}$ has the larger projection onto
$\myvec{r}_{\mathrm{trg}}$ and thus yields $\tau_{2} \equiv
\braket{\op{\rho}_{2}, \op{\rho}^{\mathrm{trg}}} > \braket{\op{\rho}_{1},
\op{\rho}^{\mathrm{trg}}} \equiv \tau_{1}$. This contradicts the expectation
that $\op{\rho}_{2}$ should be further away from $\op{\rho}^{\mathrm{trg}}$ than
$\op{\rho}_{1}$---a fact that is evidently not captured by the
functionals~\eqref{eq:oct:F_re_sm}. Moreover, the simple geometric picture of
Fig.~\ref{fig:sphere} demonstrates that the state maximizing $\tau$ is always
the pure state $\op{\rho}_{\mathrm{max}}$ with $\myvec{r}_{\mathrm{max}}
\parallel \myvec{r}_{\mathrm{trg}}$ and $\theta=0$. This readily explains the
optimization results of Fig.~\ref{fig:toy}.

\subsection{A Bloch vector-based functional for mixed state targets}
\label{subsec:funcs_new}
Before inspecting specific measures for the closeness of two mixed states to
replace an overlap-based functional~\eqref{eq:oct:F_re_sm}, we summarize the
desirable properties that a distance measure should satisfy for use in OCT\@.
Let $S_{\mathcal{H}}$ be the space of density matrices over the Hilbert space
$\mathcal{H}$. A function $D$, which quantifies the distance of two states
$\op{\rho}_{1}, \op{\rho}_{2} \in S_{\mathcal{H}}$, should fulfill
\begin{enumerate}
  \item $\forall \op{\rho}_{1}, \op{\rho}_{2} \in S_{\mathcal{H}}:
    D\left(\op{\rho}_{1}, \op{\rho}_{2}\right) \in \mathbb{R}$\,,
  \item $D\left(\op{\rho}_{1}, \op{\rho}_{2}\right) =
    \underset{\op{\rho}_{1}', \op{\rho}_{2}' \in S_{\mathcal{H}}}{\mathrm{inf}}
    D\left(\op{\rho}_{1}', \op{\rho}_{2}'\right)
    \quad \Leftrightarrow \quad
    \op{\rho}_{1} = \op{\rho}_{2}$\,.
\end{enumerate}
These two conditions provide the minimal framework for suitable state-to-state
optimization functionals. The first property ensures that an order relation can
be established. This is essential since it allows the optimization algorithm to
quantify improvement in terms of a decrease of the function value. The second
property guarantees that the minimal value identifies the desired state
uniquely~\footnote{A proper metric on a Hilbert space would furthermore require
symmetry and the triangular inequality. For use of $D$ as optimization
functional, this is not necessary.}.

Evidently, the second property is not met by the overlap-based functionals
$D_{\mathrm{re}}$ and $D_{\mathrm{sm}}$. However, there exist various distance
measures satisfying the two properties~\cite{Gilchrist.PRA.71.062310,
Dajka.PRA.84.032120}, for instance, the trace distance~\cite{NielsenChuang},
\begin{align} \label{eq:tr_dist}
  D_{\mathrm{tr}}\left(\op{\rho}_{1}, \op{\rho}_{2}\right)
  =
  \frac{1}{2} \opnorm{\op{\rho}_{1} - \op{\rho}_{2}}
  _{\mathrm{tr}}\,,
  \;
  \opnorm{\op{\rho}}_{\mathrm{tr}}
  =
  \mathrm{Tr}\left\{\sqrt{\op{\rho}^{\dagger} \op{\rho}}\right\}\,,
\end{align}
the Bures distance, based on the Uhlmann
fidelity~\cite{Uhlmann.RepMathPhys.9.273, Jozsa.JModOpt.41.2315},
\begin{align} \label{eq:bures_dist}
  D_{\mathrm{Bures}}\left(\op{\rho}_{1}, \op{\rho}_{2}\right)
  =
  \sqrt{%
    1 - \mathrm{Tr}\left\{\sqrt{%
        \sqrt{\op{\rho}_{1}} \op{\rho}_{2} \sqrt{\op{\rho}_{1}}
    }\right\}
  },
\end{align}
the Hellinger distance~\cite{Luo.PRA.69.032106},
\begin{align} \label{eq:hellinger_dist}
  D_{\mathrm{Hellinger}}\left(\op{\rho}_{1}, \op{\rho}_{2}\right)
  =
  \sqrt{%
    1 - \mathrm{Tr}\left\{\sqrt{\op{\rho}_{1}} \sqrt{\op{\rho}_{2}}\right\}
  },
\end{align}
the Jensen-Shannon divergence~\cite{Majtey.PRA.72.052310},
\begin{align} \label{eq:JS_div}
  D_{\mathrm{JS}}\left(\op{\rho}_{1}, \op{\rho}_{2}\right)
  =
  \sqrt{%
    E\left(\frac{\op{\rho}_{1} + \op{\rho}_{2}}{2}\right)
    -
    \frac{1}{2} E\left(\op{\rho}_{1}\right)
    -
    \frac{1}{2} E\left(\op{\rho}_{2}\right)
  },
\end{align}
with $E(\op{\rho}) = \mathrm{Tr}\{\op{\rho}\,
\mathrm{ln}\left(\op{\rho}\right)\}$ the von Neumann
entropy~\cite{Bennet.PRA.53.2046}, and the Hilbert-Schmidt
distance~\cite{Dodonov.JModOPt.47.633},
\begin{align} \label{eq:hs_dist}
  D_{\mathrm{HS}}\left(\op{\rho}_{1}, \op{\rho}_{2}\right)
  &=
  \frac{1}{2}
  \mathrm{Tr}\left\{%
    \left(\op{\rho}_{1} - \op{\rho}_{2}\right)^{2}
  \right\},
\end{align}
to name just a few. Note that $D_{\mathrm{HS}}$ appears in a slightly modified
version to match the definition in Ref.~\cite{Xu.JChemPhys.120.6600}, and some
of the other measures have been adapted to satisfy $D \in [0,1]$. Although
measures~\eqref{eq:tr_dist}-\eqref{eq:hs_dist} fulfill the two properties,
there is still one caveat left: With regard to gradient-based optimization
almost all of them suffer from the fact that no closed analytical expression for
their derivatives with respect to $\op{\rho}_{1}$ or $\op{\rho}_{2}$ exists.
When resorting to numerical evaluation of the derivatives, the required spectral
decomposition results in accuracy problems due to the presence of square roots,
respectively logarithms, of the state. This problem becomes particularly severe
in the common case that several eigenvalues of the density matrix are close to
zero, rendering most of the aforementioned functionals impractical for
application in gradient-based optimization.

A notable exception is the Hilbert-Schmidt distance~\eqref{eq:hs_dist}, which
therefore has already 
found use in gradient-based OCT~\cite{Xu.JChemPhys.120.6600}.
Motivated by the simple geometric picture of state mismatch, cf.
Fig.~\ref{fig:sphere}, one can also easily understand, why $D_{\mathrm{HS}}$, in
contrast to $D_{\mathrm{re}}$ or $D_{\mathrm{sm}}$, is reliable. In terms of
Bloch vectors, it reads
\begin{align}
  D_{\mathrm{HS}}\left(\op{\rho}_{1}, \op{\rho}_{2}\right)
  &=
  \frac{1}{2} \lvert \myvec{r}_{1} - \myvec{r}_{2} \rvert^{2},
\end{align}
where $D_{\mathrm{HS}}=0$ is only attainable in case of identical Bloch vectors.

While the Hilbert-Schmidt distance mixes ``angle'' and ``length'' mismatch in
a single term, one might wonder whether splitting up both contributions into two
separate terms, say $D_{\mathrm{angle}}$ and $D_{\mathrm{length}}$, allows for
a more targeted optimization. We therefore define
\begin{align} \label{eq:F_new}
  D_{\mathrm{split}} \left(\op{\rho}_{1}, \op{\rho}_{2}\right)
  &=
  \alpha_{1} D_{\mathrm{angle}} \left(\op{\rho}_{1}, \op{\rho}_{2}\right)
  +
  \alpha_{2} D_{\mathrm{length}} \left(\op{\rho}_{1}, \op{\rho}_{2}\right),
\end{align}
where $\alpha_{1}, \alpha_{2} \geq 0$ are numerical parameters that allow to
weight the contributions individually, and
\begin{subequations}
  \begin{align}
    D_{\mathrm{angle}} \left(\op{\rho}_{1}, \op{\rho}_{2}\right)
    &=
    \frac{1}{\pi^{2}}
    \arccos^{2}\left(
      \frac{d_{12}}{\sqrt{d_{11} d_{22}}}
    \right)\,,
    \\
    D_{\mathrm{length}} \left(\op{\rho}_{1}, \op{\rho}_{2}\right)
    &=
    \frac{N}{N-1}
    \left(
      \sqrt{d_{11}} - \sqrt{d_{22}}
    \right)^{2}\,,
  \end{align}
\end{subequations}
where we have used $d_{ij} \equiv \Braket{\op{\rho}_{i}, \op{\rho}_{j}} - 1/N
= \myvec{r}_{i} \cdot \myvec{r}_{j}$. In the Bloch representation, we find the
simpler form
\begin{subequations}
  \begin{align}
    D_{\mathrm{angle}} \left(\op{\rho}_{1}, \op{\rho}_{2}\right)
    &=
    \frac{\theta^{2}}{\pi^{2}}\,,
    \\
    D_{\mathrm{length}} \left(\op{\rho}_{1}, \op{\rho}_{2}\right)
    &=
    \frac{N}{N-1} \Big(
      \left|\myvec{r}_{1}\right|
      -
      \left|\myvec{r}_{2}\right|
    \Big)^{2}\,,
  \end{align}
\end{subequations}
from which it is clear that both terms quantify angle and length mismatch
individually. Measure~\eqref{eq:F_new} fulfills all required properties and can
easily be derived with respect to $\op{\rho}_{1}$ and $\op{\rho}_{2}$. The
derivatives read~\footnote{The derivation can formally be calculated by
expanding the states in any basis. The gradient is then with respect to the
coefficients in this basis, see e.g. Ref.~\cite{Palao.PRA.68.062308}.}
\begin{subequations}
  \begin{align}
    \nabla_{\op{\rho}_{1}}
    D_{\mathrm{angle}} \left(\op{\rho}_{1}, \op{\rho}_{2}\right)
    &=
    - \frac{2}{\pi^{2}}
    \arccos\left(
      \frac{d_{12}}{\sqrt{d_{11} d_{22}}}
    \right)
    \notag \\
    &\quad \times
    \frac{%
      \op{\rho}_{2} - \frac{d_{12}}{d_{11}} \op{\rho}_{1}
    }{%
      \sqrt{d_{11} d_{22} - d_{12}^{2}}
    }\,,
    \\
    \nabla_{\op{\rho}_{1}}
    D_{\mathrm{length}} \left(\op{\rho}_{1}, \op{\rho}_{2}\right)
    &=
    2 \frac{N}{N-1}
    \frac{\sqrt{d_{11}} - \sqrt{d_{22}}}{\sqrt{d_{11}}}
    \op{\rho}_{1}\,.
  \end{align}
\end{subequations}

\section{Application: Generation of Mixed State Squeezing} \label{sec:app}
\subsection{Model and Control Problem}
\label{subsec:model}
We follow Ref.~\cite{Kronwald.PRA.88.063833} and consider a mode of an optical
cavity and one of a mechanical resonator, coupled via radiation pressure. The
optical cavity is driven by two lasers at the mechanical sidebands,
$\omega_{\pm} = \omega_{\mathrm{cav}} \pm \Omega$, where $\omega_{\mathrm{cav}}$
and $\Omega$ are the frequencies of cavity and mechanical resonator,
respectively. In the linearized regime, the Hamiltonian describing the joint
system of cavity and resonator reads~\cite{Kronwald.PRA.88.063833}
\begin{equation} \label{eq:app:H}
  \begin{aligned}
    \op{H}
    &=
    - \hbar \op{d}^{\dagger} \left(
      G_{+} \op{b}^{\dagger} + G_{-} \op{b}
    \right)
    + \mathrm{H.c.}
    \\
    &\quad
    - \hbar \op{d}^{\dagger} \left(
      G_{+} \op{b} e^{- 2 \im \Omega t}
      + G_{-} \op{b}^{\dagger} e^{2 \im \Omega t}
    \right)
    + \mathrm{H.c.}\,,
  \end{aligned}
\end{equation}
where $\op{d}$ ($\op{b}$) are the annihilation operators for photons (phonons).
$G_{+}$ ($G_{-}$) are effective optomechanical coupling rates, given by the
optomechanical coupling constant times the amplitude of the lasers driving the
blue (red) sideband of the cavity. To account for decay, we use the quantum
optical master equation~\cite{Breuer.book},
\begin{align}
  \label{eq:app:LvN}
  \im \hbar \frac{\dd}{\dd t} \op{\rho}(t)
  &=
  \lop{L} \op{\rho}(t)
  \\
  &=
  \left[\op{H}, \op{\rho}(t)\right]
  +
  \im \hbar \sum_{l=1}^3 \Big(%
    \op{L}_{l} \op{\rho}(t) \op{L}_{l}^{\dagger}
    - \frac{1}{2} \big\{%
      \op{L}_{l}^{\dagger} \op{L}_{l}, \op{\rho}(t)
    \big\}
  \Big)\,,
  \notag
\end{align}
to describe the system's dynamics. The Lindblad operators are given by
\begin{subequations}
  \begin{align}
    \label{eq:app:L1}
    \op{L}_{1}
    &=
    \sqrt{\kappa}\, \op{d}\,,
    \\
    \label{eq:app:L2}
    \op{L}_{2}
    &=
    \sqrt{\Gamma_{M} \left(n_{\mathrm{th}} + 1\right)}\, \op{b}\,,
    \\
    \label{eq:app:L3}
    \op{L}_{3}
    &=
    \sqrt{\Gamma_{M} n_{\mathrm{th}}}\, \op{b}^{\dagger}
  \end{align}
\end{subequations}
with $\kappa$ and $\Gamma_{M}$ the photon and phonon decay rates, respectively,
and $n_{\mathrm{th}}$ describing the thermal occupancy of the mechanical
resonator~\cite{Kronwald.PRA.88.063833}.

Equation~\eqref{eq:app:LvN} models the driven dissipative time evolution with
steady state $\op{\rho}^{\,\mathrm{th}}$. In other words, the optomechanical
system will end up in $\op{\rho}^{\,\mathrm{th}}$, independent of the initial
state $\op{\rho}(0)$, if one waits sufficiently long, i.e., $\op{\rho}(0)
\rightarrow \op{\rho}^{\,\mathrm{th}}$ for $t \rightarrow \infty$. The reduced
steady state of the resonator alone is obtained by taking the partial trace over
the cavity mode, $\op{\rho}^{\,\mathrm{th}}_{\mathrm{res}}
= \mathrm{Tr}_{\mathrm{cav}}\left\{\op{\rho}^{\,\mathrm{th}}\right\}$. An
appropriate choice of coupling $G_{-}$ and relative strength $G_{+}/G_{-} < 1$
results in squeezed thermal steady states of the
resonator~\cite{Kronwald.PRA.88.063833}, where the squeezing strength is
quantified by the expectation value $\langle \op{X}_{1}^{2} \rangle$ of the
mechanical quadrature, $\op{X}_{1} = (\op{b} + \op{b}^{\dagger})/\sqrt{2}$. It
was found~\cite{Kronwald.PRA.88.063833} that larger squeezing of
$\op{\rho}_{\mathrm{res}}^{\mathrm{th}}$ is usually accompanied by lower purity
and vice versa. The generation of strongly squeezed states comes thus at the
expense of lower purity.

Note that the purity of $\op{\rho}^{\mathrm{th}}_{\mathrm{res}}$, as well as
that of $\op{\rho}^{\mathrm{th}}$ is in general determined by $\kappa,
\Gamma_{M}$ and $n_{\mathrm{th}}$, in addition to $G_{+}$ and $G_{-}$. In cavity
optomechanics, the joint effect of these parameters is captured by the
cooperativity, $\mathcal{C} = 4 G_{-}^{2}/(\kappa \Gamma_{M})$. It serves as
figure of merit for any optomechanical system, quantifying the exchange of
photons and phonons, i.e., the coupling between optical cavity and mechanical
resonator~\cite{Yuan.NatCommun.6.8491, Aspelmeyer.RevModPhys.86.1391}.

If the laser drives operate continuously, the time $T$ it takes to reach
$\op{\rho}^{\,\mathrm{th}}$ with sufficient accuracy is essentially determined
by the cooperativity $\mathcal{C}$ and the optomechanical coupling rates
$G_{+}$, $G_{-}$. Assuming cavity and resonator to be initially in thermal
equilibrium, we may ask whether it is possible to accelerate the approach
of the steady state by suitably shaping the drives. To this end, we consider
time-dependent driving strengths of the blue and red sideband tones. This
results in time-dependent effective coupling rates $G_{-}(t)$ and $G_{+}(t)$. We
will use optimal control theory as outlined in Sec.~\ref{sec:oct} to determine
shapes of $G_{-}(t)$ and $G_{+}(t)$ that allow for a faster approach to the
steady state compared to the constant drives of
Ref.~\cite{Kronwald.PRA.88.063833}.

\begin{figure}[tb]
  \centering
  \includegraphics{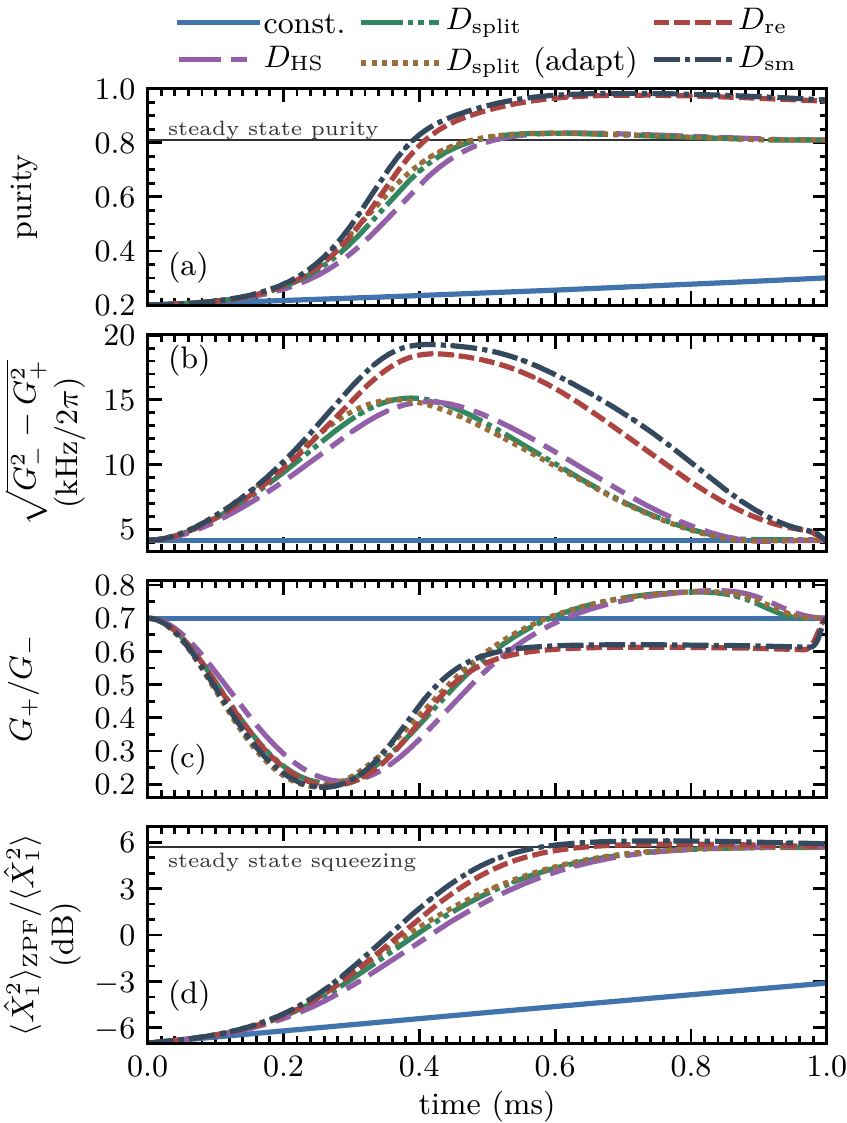}
  \caption{%
    Comparison of the time evolution obtained with time-constant (blue solid
    lines) and optimized drives, using the target functionals indicated in the
    legend:
    (a)
    joint state purity,
    (b) and (c)
    optimized effective cooling rate $(G_{-}^{2}(t) - G_{+}^{2}(t))^{1/2}$ and
    squeezing rate $G_{+}(t)/G_{-}(t)$, respectively,
    (d)
    squeezing strength $\langle \op{X}_{1}^{2} \rangle$ compared to the
    zero-point fluctuations $\langle \op{X}_{1}^{2} \rangle_{\mathrm{ZPF}}$ in
    dB, i.e., $10 \cdot \log_{10}\{\langle \op{X}_{1}^{2} \rangle_{\mathrm{ZPF}}
    / \langle \op{X}_{1}^{2} \rangle\}$.
    ``$D_{\mathrm{split}}$ (adapt)'' represents an optimization using an
    adaptive choice for the weights of angle and length, cf.\
    Eq.~\eqref{eq:app:adapt}.
  }
  \label{fig:squeeze_dyn}
\end{figure}

\subsection{Speeding up the approach of the steady state}
\label{subsec:results}
The assumption of thermal equilibrium initially corresponds, for the cavity, to
the ground state~\footnote{For simplicity we take the ground state as the
initial state of the cavity.  However, the optimization works for different
coherent initial cavity states as well, since the general timescale for
approaching the steady state remains the same.}, $\op{\rho}_{\mathrm{cav}}(0)
= \Ket{0}\Bra{0}$, whereas the initial state of the resonator is characterized
by the thermal occupancy $n_{\mathrm{th}}$, for which we choose $n_{\mathrm{th}}
= 2$~\footnote{Note that we neglect initial correlations between cavity and
resonator modes}. Cavity and resonator decay rate are taken from the experiment
reported in Ref.~\cite{Wollman.Science.349.952}, i.e., $\kappa/2\pi
= \SI{450}{\kilo\hertz}$, $\Gamma_{M}/2\pi = \SI{3}{\hertz}$. The target state
for the optimization is given by the state obtained with the time-continuous
protocol of Ref.~\cite{Kronwald.PRA.88.063833} after $\SI{15}{\milli\second}$
which is virtually identical to the steady state. For the given parameters, the
squeezing strength in the steady state, $\langle \op{X}_{1}^{2}
\rangle_{\mathrm{ZPF}} / \langle \op{X}_{1}^{2} \rangle$, amounts to
approximately $5.7$\,dB which is beyond the 3\,dB limit. Constant values for
$G_{-}$ and $G_{+}$, cf. Ref.~\cite{Kronwald.PRA.88.063833}, are taken as
a guess pulse for starting the iterative optimization. In detail, we choose
$G_{+}$ and $G_{-}$ such that $\mathcal{C} = 100$ and $G_{+}/G_{-} = 0.7$, since
this balances well squeezing and mixedness of the associated steady state. In
the calculations, $N_{\mathrm{res}} = 40$ and $N_{\mathrm{cav}} = 4$ levels for
resonator and cavity mode turn out to be sufficient to prevent reflection due to
the finite Hilbert space size. Fast oscillating terms in Eq.~\eqref{eq:app:H}
have been neglected, which is, given our choice of $\mathcal{C}$, in accordance
with Ref.~\cite{Kronwald.PRA.88.063833}.

\begin{figure}[tb]
  \centering
  \includegraphics{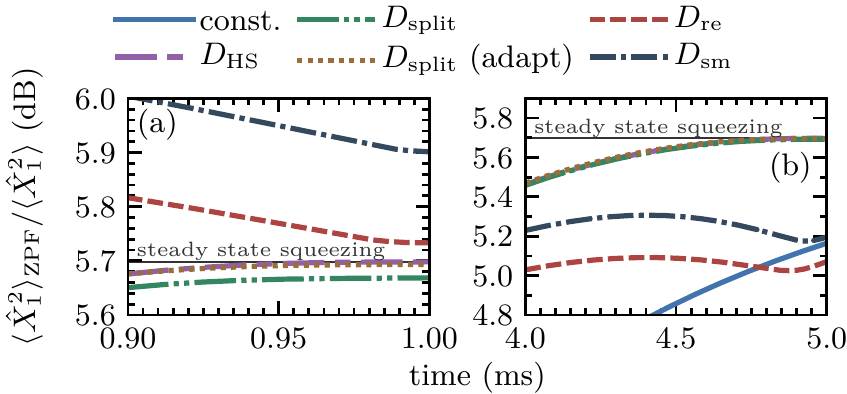}
  \caption{%
    (a) The relevant part of the squeezing dynamics as shown in
    Fig.~\ref{fig:squeeze_dyn}(d). (b) A similar dynamics as in (a) but for an
    optimization with final time $\SI{5}{\milli\second}$ (optimized fields and
    dynamics not shown).
  }
  \label{fig:squeeze_inset}
\end{figure}

Figure~\ref{fig:squeeze_dyn} compares the dynamics of the time-continuous
protocol of Ref.~\cite{Kronwald.PRA.88.063833} (blue solid lines) to those
induced by optimized drives, using several target functionals and a total time
of $\SI{1}{\milli\second}$. The joint state purity and resonator squeezing are
analyzed in Fig.~\ref{fig:squeeze_dyn}(a) and (d), respectively. Moreover,
Fig.~\ref{fig:squeeze_dyn}(b) shows the difference $(G_{-}^{2}(t)
- G_{+}^{2}(t))^{1/2}$, which determines an effective cooling rate into the
squeezed state while Fig.~\ref{fig:squeeze_dyn}(c) shows the ratio
$G_{+}(t)/G_{-}(t)$, an effective rate determining the squeezing strength of the
final steady state, cf. Ref.~\cite{Kronwald.PRA.88.063833}.  Pulses optimized
using $D_{\mathrm{HS}}$ or $D_{\mathrm{split}}$ result in an acceleration of the
thermalization process, cf.\ the blue solid vs.\ purple double-dashed, brown
dotted and green dashed-double dotted lines in Fig.~\ref{fig:squeeze_dyn}(a).
These lines all converge to the proper joint state purity. Similarly, the
resonator squeezing reaches the desired value for the corresponding curves in
Fig.~\ref{fig:squeeze_dyn}(d) and does so significantly faster for all optimized
pulses.

Inspection of Fig.~\ref{fig:squeeze_dyn}(b) and (c) allows us to unravel the
control strategy. It consists, independently of the target functional, in an
increase of the effective cooling rate $(G_{-}^{2}(t) - G_{+}^{2}(t))^{1/2}$ in
order to speed up the cooling into the (squeezed) steady state. In general,
ramping up $G_{-}$ and $G_{+}$ will always accelerate the coherent part of the
dynamics, since the norm of the Hamiltonian~\eqref{eq:app:H} determines the
timescale of the system's coherent dynamics. However, ramping up the coupling
also changes the steady state of the driven dissipative dynamics. Thus, the
increase of $(G_{-}^{2}(t) - G_{+}^{2}(t))^{1/2}$, which is in our case in fact
achieved by increasing both $G_{-}(t)$ and $G_{+}(t)$, needs to be balanced by
a modulation of $G_{+}(t)/G_{-}(t)$ to ensure steering the system towards the
correct target state. Interestingly, the optimizations with both
$D_{\mathrm{HS}}$ (purple double-dashed lines) and $D_{\mathrm{split}}$ (green
dashed-double dotted and brown dotted lines) find almost identical control
fields. This is not guaranteed due to non-uniqueness of the control solution in
most cases and indicates that we explore comparable optimization
landscapes~\cite{TibbettsJCP13, RivielloPRA15} despite the different
functionals.

Note that the optimized control fields of Fig.~\ref{fig:squeeze_dyn}(b-c) only
require a slow modulation of the drive amplitudes while keeping their
frequencies constant. This makes them experimentally feasible with existing
technology---such slow modulations can easily be realized by arbitrary waveform
generators allowing for amplitude modulations on timescales down to
sub-nanoseconds~\cite{Rogers.RSI.82.073107} or even significantly more complex
pulse shapes, see e.g. Ref.~\cite{HaeberlePRL13} for one example.

Another concern that often arises in the context of experimental feasibility of
optimal control protocols is robustness with respect to noise in the controls.
We therefore examine whether our optimized drives are robust with respect to
amplitude noise. To this end, we apply $0.2\%, 0.5\%$ and $1.0\%$ constant noise
to the optimized field shapes~\footnote{We use an optimization with total time
$T=\SI{5}{\milli\second}$ as benchmark.} by rescaling the field amplitudes
accordingly. We obtain for the final trace distance $D_{\mathrm{tr}}$ with
respect to the targeted squeezed steady state an average of $3.7 \cdot 10^{-4}$,
$9.2 \cdot 10^{-4}$ and $1.8 \cdot 10^{-3}$, respectively. This needs to be
compared to $D_{\mathrm{tr}} = 4.8 \cdot 10^{−6}$ for the original optimized
fields and to $D_{\mathrm{tr}} = 2.6 \cdot 10^{−2}$ which one would obtain under
the evolution with constant drives up to that point in time. The increase of the
absolute error, of the order of $10^{-4}-10^{-3}$, is not surprising giving the
order of the noise which is $10^{-3}-10^{-2}$.

Figure~\ref{fig:squeeze_inset}(a) provides a closer look at the asymptotic
squeezing dynamics of Fig.~\ref{fig:squeeze_dyn}(d), showing that only pulses
optimized with $D_{\mathrm{HS}}$ and $D_{\mathrm{split}}$ reach the correct
squeezing at final time, cf.\ the purple double-dashed and brown dotted lines.
In contrast, the fields optimized with $D_{\mathrm{re}}$ and $D_{\mathrm{sm}}$
(red dashed and dark blue dot-dashed lines in Fig.~\ref{fig:squeeze_dyn}) fail
to steer the system towards a state with the correct purity. Instead, they act
in order to increase the purity at final time $T$ as much as possible, failing
to reach, however, completely pure states which are not attainable due to the
finite temperature ($n_{\mathrm{th}} > 0$). Figures~\ref{fig:squeeze_dyn}
and~\ref{fig:squeeze_inset}(a) thus illustrate once more that the target
functionals $D_{\mathrm{re}}$ and $D_{\mathrm{sm}}$ should not be used for
non-pure target states.

Interestingly, the dynamics shown in Fig.~\ref{fig:squeeze_inset}(a) all results
in a comparable squeezing with the final values $\langle \op{X}_{1}^{2}
\rangle_{\mathrm{ZPF}}/\langle \op{X}_{1}^{2}\rangle$ obtained with
$D_{\mathrm{re}}$ and $D_{\mathrm{sm}}$ even beyond the intended steady state
squeezing of roughly 5.7\,dB, cf.\ the red dashed and dark blue dot-dashed lines in
Fig.~\ref{fig:squeeze_inset}(a). This indicates a larger squeezing to be
possible than the one set by the steady state with its corresponding tradeoff
between squeezing strength and purity. The apparently ``good'' optimization
results with respect to the final-state squeezing obtained with
$D_{\mathrm{re}}$ and $D_{\mathrm{sm}}$ in Fig.~\ref{fig:squeeze_inset}(a) can
be explained by the fact that squeezing of any state is mainly determined by its
direction on the generalized Bloch sphere. Here, the optimization benefits from
the fact that $D_{\mathrm{re}}$ and $D_{\mathrm{sm}}$ try to match the final
state directions. However, this is not always the case, as illustrated in
Fig.~\ref{fig:squeeze_inset}(b) showing the squeezing dynamics for a similar
optimization as in Fig.~\ref{fig:squeeze_dyn} but with a final time of
$\SI{5}{\milli\second}$. Note that for the constant protocol of
Ref.~\cite{Kronwald.PRA.88.063833} the optimal relation $G_{+}/G_{-}$ of driving
strengths which maximizes the squeezing strength can be estimated by
$n_{\mathrm{th}}$ and the cooperativity $\mathcal{C}$. However, this estimation
is no longer easily possible for shaped driving fields since they give rise to
time-dependent cooperativities $\mathcal{C}(t)$.

\begin{figure}[tb]
  \centering
  \includegraphics{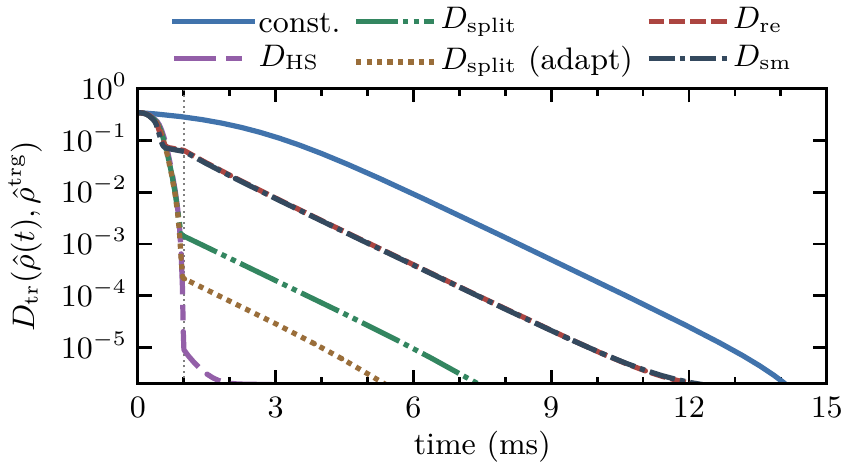}
  \caption{%
    Dynamics of the trace distance $D_{\mathrm{tr}}$, cf.
    Eq.~\eqref{eq:tr_dist}, under the guess and optimized fields of
    Fig.~\ref{fig:squeeze_dyn}. Beyond $T=\SI{1}{\milli\second}$ (indicated by
    vertical line) all optimized fields are extended by the constant fields of
    the original, time-independent protocol.
  }
  \label{fig:dyn_measure}
\end{figure}
If the state preparation errors obtained with the optimized fields after
$\SI{1}{\milli\second}$ 
are not yet sufficient, it
should be possible to continue approaching the steady state using the original
protocol~\cite{Kronwald.PRA.88.063833} of constant drives. This is examined in
Fig.~\ref{fig:dyn_measure} which shows the evolution of the trace distance
$D_{\mathrm{tr}}$, cf. Eq.~\eqref{eq:tr_dist}, under constant drives and
optimized fields from Fig.~\ref{fig:squeeze_dyn}(b) and (c), switched back to
constant drives at $T = \SI{1}{\milli\second}$. $D_{\mathrm{tr}}$ continues to
decrease for times larger than the switching time, i.e., the final time used in
the optimization. A monotonous decrease of $D_{\mathrm{tr}}$ across the
switching time, as observed in Fig.~\ref{fig:dyn_measure}, is expected for the
fields optimized with $D_{\mathrm{split}}$ and $D_{\mathrm{HS}}$. It does not
need to be the case, however, for the fields optimized with $D_{\mathrm{re}}$ or
$D_{\mathrm{sm}}$. Here, the state at the switching time, although already
closer to the target state than with constant driving, is still comparatively
far from the steady state. Nevertheless, upon subsequent propagation with
constant drives, $D_{\mathrm{tr}}$ is further improved in all cases.
Figure~\ref{fig:dyn_measure} thus provides another illustration of the speed up
in preparing the squeezed steady state.

\begin{figure}[tb]
  \centering
  \includegraphics{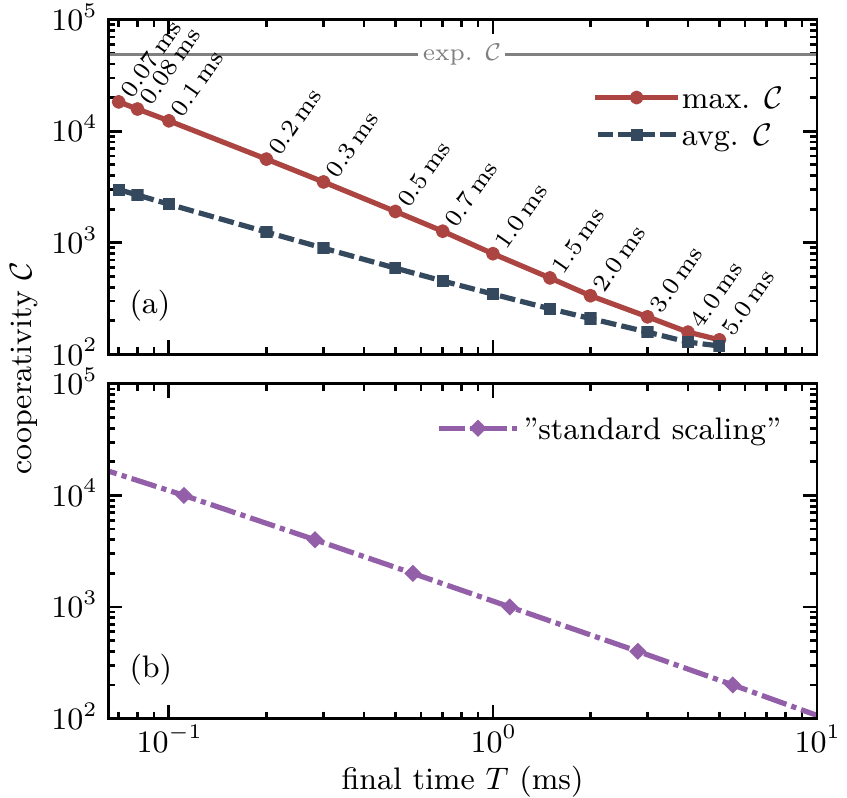}
  \caption{%
    (a)
    Peak and average cooperativity $\mathcal{C}$, calculated from the optimized
    fields $G_{-}(t)$, required to achieve a state preparation error of at most
    $D_{\mathrm{tr}} < 10^{-4}$, as function of the total optimization time $T$
    (employing $D_{\mathrm{HS}}$ for all optimizations). The horizontal line
    indicates the static cooperativity used in the experiment reported in
    Ref.~\cite{Wollman.Science.349.952}.
    (b)
    Minimal time against required cooperativity to reach a steady state with
    $D_{\mathrm{tr}} < 10^{-4}$ for the original, time-constant protocol of
    Ref.~\cite{Kronwald.PRA.88.063833}. Note that each point in (b) corresponds
    to a \emph{different} steady state while all points in (a) correspond to the
    \emph{same} steady state.
  }
  \label{fig:coop}
\end{figure}

Finally, Fig.~\ref{fig:coop} answers the question by how much the approach of
the steady state can be accelerated. The price for speed-up is cooperativity,
or, in other words, laser intensity, as illustrated by Fig.~\ref{fig:coop}(a).
It shows the peak and average cooperativity $\mathcal{C}$, determined by the
optimized field $G_{-}(t)$, as a function of the total optimization time $T$.
Given an experimental bound on the cooperativity $\mathcal{C}$, one can thus
easily determine the required time $T$ to reach the target state. Taking the
experimental value of the cooperativity reported in
Ref.~\cite{Wollman.Science.349.952}, we find a speedup of at least two orders of
magnitude, see Fig.~\ref{fig:coop}(a), compared to the original protocol
employing constant drives~\cite{Kronwald.PRA.88.063833}. Conversely, fixing
a certain duration $T$ determines the required cooperativity or laser power.
Durations as short as $T=\SI{0.07}{\milli\second}$ are feasible, while the state
preparation errors are still sufficiently small with $D_{\mathrm{tr}} < 10^{-4}$
for all points in Fig.~\ref{fig:coop}(a). Moreover, the optimized pulse shapes
corresponding to the data from Fig.~\ref{fig:coop}(a) all look quite similar to
the ones presented in Fig.~\ref{fig:squeeze_dyn}(b) and~(c).

Both peak and average cooperativity of the optimized field increase with
decreasing duration, as one would expect for reaching the same target in less
time. We observe an almost perfect power law dependence of the cooperativity
$\mathcal{C}$ as a function of the duration $T$ in Fig.~\ref{fig:coop}(a). This
power law should be compared to the intrinsic scaling of the system due to its
non-linearity which is shown in Fig.~\ref{fig:coop}(b). Note that each point in
Fig.~\ref{fig:coop}(b) corresponds to a different steady state while all points
in Fig.~\ref{fig:coop}(a) correspond to the same steady state, namely the one
used as benchmark in Fig.~\ref{fig:squeeze_dyn} after $\SI{15}{\milli\second}$.
The similar scaling observed in Fig.~\ref{fig:coop}(a) and (b) thus indicates
the system non-linearity to be the defining feature even in the case of
time-dependent and optimized drives.

Note that the short fields with cooperativities $\mathcal{C} \sim 10^{4}$
approach the regime where the rotating wave approximation starts to break
down~\cite{Kronwald.PRA.88.063833}. Hence, we chose not to examine shorter,
respectively stronger fields in Fig.~\ref{fig:coop}.

\subsection{Performance of optimization functionals}
\label{subsec:results:perf}

\begin{figure}[tb]
  \centering
  \includegraphics{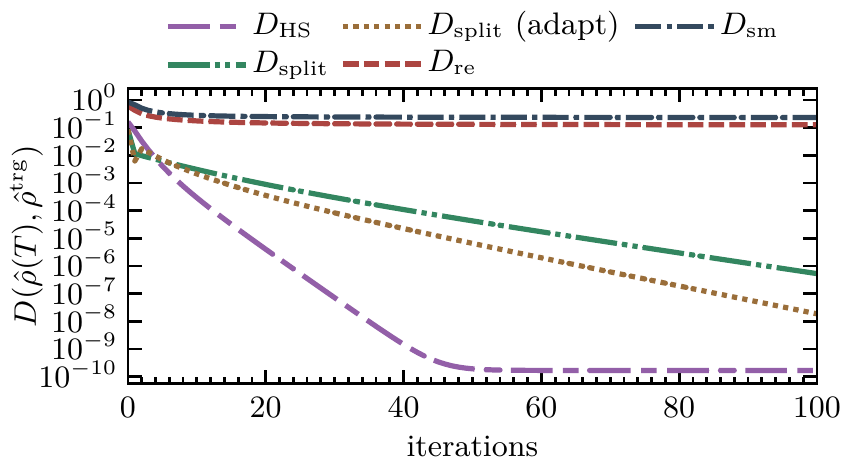}
  \caption{%
    Convergence behavior of optimization algorithm for the target functionals
    used in Fig.~\ref{fig:squeeze_dyn}. The weighting of the two terms in
    $D_{\mathrm{split}}$ was chosen as $\alpha_{1} = \alpha_{2} = 1/2$, cf.
    Eq.~\eqref{eq:F_new}, for the green dashed double-dotted line, while it was
    adapted after each iteration for the brown dotted line (see text).
  }
  \label{fig:squeeze_iters}
\end{figure}

The convergence behavior of the optimization algorithm for the various target
functionals used in Fig.~\ref{fig:squeeze_dyn} is inspected in
Fig.~\ref{fig:squeeze_iters}. The functional value of $D_{\mathrm{re}}$ and
$D_{\mathrm{sm}}$ rapidly approaches a plateau which indicates that the
optimization got stuck and no improvement with respect to the guess pulses could
be realized. In contrast, optimizations with $D_{\mathrm{HS}}$ (purple
double-dashed line) and $D_{\mathrm{split}}$ (green dashed-double dotted and
brown dotted lines) show an improvement of several orders of magnitude. For the
optimizations with $D_{\mathrm{split}}$, we have used two different variants.
For the green dashed double-dotted line constant, equivalent weights
$\alpha_{1}, \alpha_{2}$, cf. Eq.~\eqref{eq:F_new}, have been used, while for
the brown dotted line we have employed an automated update scheme for the
weights after each iteration. In the latter case, we have adjusted the weights
for the next iteration $i+1$,
\begin{align} \label{eq:app:adapt}
  \alpha_{1}^{(i+1)}
  =
  \frac{%
    D_{\mathrm{angle}}^{(i)}
  }{%
    D_{\mathrm{angle}}^{(i)} + D_{\mathrm{radius}}^{(i)
  }}\,,
  \quad
  \alpha_{2}^{(i+1)}
  =
  \frac{%
    D_{\mathrm{radius}}^{(i)}
  }{%
    D_{\mathrm{angle}}^{(i)} + D_{\mathrm{radius}}^{(i)
  }}\,,
\end{align}
using values of the current iteration $i$. This effectively causes the
dominating term to become preferentially minimized within the next iteration.
Although breaking strict monotonic convergence of Krotov's method over multiple
iterations, due to optimizing a different functional in each iteration, it
yields better convergence in our example. The plateau of $D_{\mathrm{HS}}$ at
$\sim 10^{-10}$, starting at iteration $\sim 50$, is not of physical origin but
caused by the propagation accuracy; choosing a finer time discretization would
probably allow the optimization to reach even smaller values.

Note that the scaling parameters $\lambda_{k}$, cf. Eq.~\eqref{eq:krotov:g},
have been individually chosen for the different functionals in all optimizations
shown in Fig.~\ref{fig:squeeze_iters}~\footnote{The $\lambda_{k}$'s are taken to
be identical for both fields, i.e., $\lambda = \lambda_{+} = \lambda_{-}$ for
$G_{-}(t)$ and $G_{+}(t)$, respectively.}. The necessity of different scalings
is readily explained by the co-states $\op{\chi}_{l}(T)$, since their norm
influences the magnitude of the field updates, via
Eq.~\eqref{eq:krotov:co_states_bound}. Due to different norms for different
functionals, the optimization parameters $\lambda_{k}$ must usually be adjusted
when switching functionals if one wants to maintain field updates of similar
magnitude.

\begin{figure}[tb]
  \centering
  \includegraphics{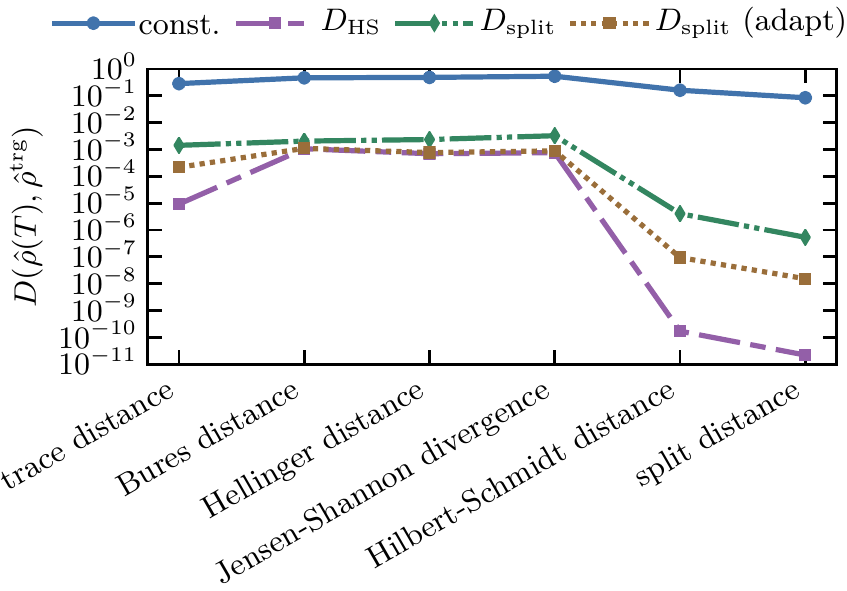}
  \caption{%
    Comparison between changes in various distance measures under optimization
    with $D_{\mathrm{HS}}$ and $D_{\mathrm{split}}$, cf.
    Fig.~\ref{fig:squeeze_dyn}.
  }
  \label{fig:measures}
\end{figure}
In the same context, one might naively conjecture from
Fig.~\ref{fig:squeeze_iters} that, because $D_{\mathrm{HS}}$ yields smaller
functional values than $D_{\mathrm{split}}$, $D_{\mathrm{HS}}$ yields better
optimization results. However, such a statement would in general be wrong. As
discussed above in Sec.~\ref{subsec:funcs_new}, the accuracy with which the
target state is reached is not uniquely assessed by a single measure. A small
value of $D_{\mathrm{HS}}$ does not necessarily imply a similarly good value for
any other distance measure. Figure~\ref{fig:measures} therefore displays the
value of several reliable distance measures for the final state obtained with
the fields optimized using $D_{\mathrm{HS}}$ and $D_{\mathrm{split}}$ and
compares it with the non-optimized protocol, i.e., constant driving (blue line).
We indeed observe that the optimization with $D_{\mathrm{HS}}$ (purple
double-dashed line) yields the smallest state-preparation errors also for all
other distance measures in Fig.~\ref{fig:measures}.

This does not need to hold in general, however, since the \emph{absolute} value
of any distance measure $D$ is not to be confused with \emph{relative} physical
closeness of two states. While the measures $D$ considered in
Fig.~\ref{fig:measures} are all known to be reliable, they assess state
mismatches differently for $D>0$. For instance, if
$
D(\op{\rho}_{2}, \op{\rho}^{\mathrm{trg}})
>
D(\op{\rho}_{1}, \op{\rho}^{\mathrm{trg}})
>
0
$
for two states $\op{\rho}_{1}, \op{\rho}_{2}$, a desired target state
$\op{\rho}^{\mathrm{trg}}$, and a measure $D$ does not imply the same to be true
for another measure $\tilde{D}$. In other words, two reliable measures can still
disagree on which of two states is closer to the target even though they both
correctly assess when a state becomes identical to the target.

For the presented problem, the performance of the Hilbert-Schmdidt distance
$D_{\mathrm{HS}}$ compared to the split-functional $D_{\mathrm{split}}$ is
slightly better, cf. Fig.~\ref{fig:squeeze_iters} and~\ref{fig:measures}.
Nevertheless, $D_{\mathrm{split}}$ contains information about angle and length
mismatch of the Bloch vectors individually and thus provides more insight into
the dominating source of mismatch which $D_{\mathrm{HS}}$ cannot provide.  While
this information was not of relevance here it could certainly be of interest for
other optimization problems.

\section{Conclusions} \label{sec:con}
We have studied how to speed up evolution towards a squeezed steady state in
a driven optomechanical system, consisting of cavity and mechanical resonator
coupled via radiation pressure. To this end, we have replaced the constant
drives of the original protocol~\cite{Kronwald.PRA.88.063833} by time-dependent
pulses and derived the corresponding pulse shapes using quantum optimal control
theory. To the best of our knowledge, our work is the first to apply quantum
optimal control to cavity optomechanics. Further potential of quantum optimal
control for this popular experimental platform is highlighted by a recent
proposal suggesting to couple the cavity additionally to a two-level system in
order to drive the mechanical oscillator into a Fock state~\cite{Bergholm2018}.

Our control solutions for accelerating the approach of a squeezed steady state
consist in increasing the effective optomechanical coupling at intermediate
times. At final time, the value of the constant coupling is resumed, ensuring
approach of the proper steady state. We find the cooperativity corresponding to
the increased optomechanical coupling due to the optimized fields to grow
polynomially with decreasing protocol duration, for both average and peak value.
Limiting the maximum cooperativity to that of the experiment reported in
Ref.~\cite{Wollman.Science.349.952}, a speed up of more than two orders of
magnitude is possible, compared to the protocol using constant drives. The
required pulse shapes correspond to simple modulations and are feasible with
current technology using e.g.\ arbitrary wave form generators. In view of using
the squeezed state, for example in quantum sensing, such a speed up will be
important to minimize the detrimental influence of decoherence.

Since the steady state balances quantum mechanical purity and resonator
squeezing, the control problem targets a non-pure state, and care must be taken
when defining the target functional. In particular, functionals based on state
overlaps fail when both states---the true state and the target state---are
mixed. A possible remedy consists in replacing the overlap by a (modified)
distance measure~\cite{Xu.JChemPhys.120.6600}. We have visualized the failure of
overlap based functionals by examining the state vectors on the Bloch sphere:
While the overlap only seeks to match the angle, a reliable figure of merit
needs to match both angle and length of the vectors. This geometric picture
provides the intuition for defining an alternative target functional, based on
matching angle and length of the Bloch vectors separately. We have successfully
employed this target functional as an alternative to a functional based on the
Hilbert Schmidt distance~\cite{Xu.JChemPhys.120.6600}, obtaining fairly similar
solutions to the control problem at hand. Moreover, we observe that optimization
with both functionals not only leads to a minimization of the respective
distance measure that is being employed but also to a reduction of any other
distance measure that can be used to assess the state preparation error.

Our results of accelerated state preparation are relevant when exploiting
squeezed states, for example in quantum sensing. Moreover, our Bloch vector
based target functional should be useful, in general, to estimate quantum speed
limits~\cite{Campaioli2018}. While the mismatch in Bloch vector angles
quantifies rotation (i.e., unitary) errors, that in Bloch vector length
estimates dissipative errors. If, for a given system, one can find an expression
for the evolution speed of Bloch vector angle and length, this would allow to
determine separate quantum speed limits for the unitary and dissipative parts of
a system's evolution. One could thus decide which of the two sets the overall
speed limit.

Our results also give rise to an interesting further question in the context of
squeezed state preparation. Incidentally, we have found fields that, while not
resulting in the correct steady state, produce larger squeezing than expected
for the steady state, with higher purity. This suggests to directly maximize the
squeezing at final time, irrespective of the state at that time, instead of
targeting a specific squeezed state as we have done here. Such an optimization
is possible by taking the expectation value of the relevant quadrature as target
functional. It would allow to examine the conditions for avoiding the trade-off
between purity and squeezing to which the steady state is subject
to~\cite{Kronwald.PRA.88.063833} and, more generally, determine the ultimate
limit of quantum mechanical squeezing.

\begin{acknowledgments}
  We thank Ronnie Kosloff and Florian Marquardt for helpful discussions.
  Financial support from the Volkswagenstiftung is gratefully acknowledged.
\end{acknowledgments}

%

\end{document}